\documentclass[traditabstract]{aa} 
\usepackage{longtable}
\usepackage{amsmath}
\usepackage{graphicx}
\usepackage{txfonts}
\usepackage{natbib}
\usepackage{verbatim}
\begin{document}
   \title{Spectroscopic characterization of a sample of metal-poor solar-type stars from the HARPS planet search program\thanks{Based on observations collected at the La Silla Parana
Observatory, ESO (Chile) with the HARPS spectrograph at the 3.6-m telescope (ESO runs ID 72.C-0488, 082.C-0212, and 085.C-0063).}}

   \subtitle{Precise spectroscopic parameters and mass estimation}

   \author{S. G. Sousa\inst{1,}\inst{2}
          \and
	  N. C. Santos\inst{1,}\inst{3,}\inst{4}
	  \and G. Israelian\inst{2,}\inst{5}
	  \and C. Lovis\inst{3}
	  \and M. Mayor\inst{3}
	  \and P. B. Silva\inst{1,}\inst{4}
	  \and
	  S. Udry\inst{3} 
          }

	  \institute{Centro de Astrof\'isica, Universidade do Porto, Rua das Estrelas, 4150-762 Porto, Portugal
	  \and Instituto de Astrof\'isica de Canarias, 38200 La Laguna, Tenerife, Spain
	  \and Geneva Observatory, Geneva University, 51 Ch. des Mailletes, 1290 Sauverny, Switzerland
	  \and Departamento de F\'isica e Astronomia, Faculdade de Ci\^encias da Universidade do Porto, Portugal
	  \and Departamento de Astrofisica, Universidade de La Laguna, E-38205 La Laguna, Tenerife, Spain}

   \date{}

 
  \abstract{Stellar metallicity strongly correlates with the presence of planets and their properties. To check for new correlations between stars and the existence of an orbiting planet, 
we determine precise stellar parameters for a sample of metal-poor
solar-type stars. This sample was observed with the HARPS spectrograph and is part of a program to search for new extrasolar planets. 

The stellar parameters were determined using an LTE analysis based on equivalent widths (EW) of iron lines and by imposing excitation and ionization equilibrium. The ARES
code was used to allow automatic and systematic derivation of the stellar parameters.

Precise stellar parameters and metallicities were obtained for 97 low metal-content stars. We also present the derived masses, luminosities, and new parallaxes estimations based on the derived parameters,
and compare our spectroscopic parameters with an infra-red flux method calibration to check the consistency of our method in metal poor stars. Both methods seems to give the same effective temperature scale.

Finally we present a new calibration for the temperature as a function of \textit{B-V} and [Fe/H]. This was obtained by adding these new metal poor stars in order to increase 
the range in metallicity for the calibration. The standard deviation of this new calibration 
is $\sim$ 50 K.
}

\authorrunning{Sousa, S. G. et al.}
\titlerunning{Spectroscopic characterization of a sample of metal-poor solar-type stars...}

\maketitle

\section{Introduction}

The discovery of exoplanets continues at a very high rate and has recently passed the 450th detection. The radial velocity 
technique gives strong input for this number, supported by the several dedicated observing programs 
that almost continually observe stellar spectra in different high-resolution 
instruments spread across the world. The HARPS spectrograph is one of the leading instruments that 
over the past five years has alone spotted more than 85 of the exoplanets now known to orbit stars other than the Sun.

All of these new discoveries are providing new clues for the formation and evolution of stars and planets. 
One of these clues is the very well known and established correlation between the metallicity of the stars
 and the presence of an orbiting giant planet \citep[][]{Gonzalez-1997, Gonzalez-2001, Santos-2001, Santos-2004b, Fischer_Valenti-2005, Udry-2006, Udry-2007b}. 
This observational correlation suggests that giant planets are 
more easily formed around metal-rich stars, supporting the core accretion idea as the main mechanism in the 
formation of giant planets \citep[][]{IdaLin-2004, Benz-2006} instead of the alternative model focused on the idea of the disk instability\citep[][]{Boss-2002}.

Although this correlation seems to be true for giant planets, it might not be so for lower mass planets. 
With the new discoveries reaching lower and lower masses, the new ``lighter systems`` are starting to reveal that these new planet-host stars present a different and wider 
metallicity distribution \citep[][]{Sousa-2008}. That can also be explained by recent models of the core accretion idea \citep[][]{Mordasini-2009}.


\begin{table*}[!t]
\centering 
\caption[]{Details of the spectral data for each star.}
\begin{scriptsize}
\begin{tabular}{lccccc|lccccc}
\hline
\hline
\noalign{\smallskip}

Name & totsp & sumsp & maxsn & minsn & sumsn & Name & totsp & sumsp & maxsn & minsn & sumsn \\
\hline

\object{HD\,102200} 	&	    7	&	    3	&	164.50	&	128.10	&	 251.75	 & \object{HD\,197890}  &	       5   &	       0   &	     0.00  &	     0.00  &	      0.00 \\
\object{HD\,104800} 	&	    6	&	    6	&	123.70	&	 70.20	&	 235.46	 & \object{HD\,199288}  &	      15   &	      11   &	   411.10  &	   272.80  &	   1137.58 \\
\object{HD\,105004} 	&	    5	&	    5	&	 81.60	&	 45.10	&	 138.91	 & \object{HD\,199289}  &	       5   &	       3   &	   140.80  &	    98.60  &	    201.48 \\
\object{HD\,107094} 	&	   12	&	    7	&	123.40	&	 95.60	&	 287.16	 & \object{HD\,199604}  &	       6   &	       4   &	   157.10  &	   144.00  &	    301.09 \\
\object{HD\,108564} 	&	    6	&	    6	&	130.70	&	 75.70	&	 246.54	 & \object{HD\,199847}  &	       7   &	       6   &	   118.40  &	    67.60  &	    210.07 \\
\object{HD\,109310} 	&	   15	&	    9	&	180.80	&	127.40	&	 434.18	 & \object{HD\,206998}  &	       6   &	       4   &	   132.00  &	    92.80  &	    224.53 \\
\object{HD\,109684} 	&	    6	&	    4	&	142.20	&	 78.00	&	 239.40	 & \object{HD\,207190}  &	       5   &	       2   &	   232.50  &	   134.30  &	    268.50 \\
\object{HD\,111515} 	&	    5	&	    2	&	193.30	&	188.60	&	 270.06	 & \object{HD\,207869}  &	      17   &	      17   &	   135.30  &	    41.20  &	    425.99 \\
\object{HD\,111777} 	&	    6	&	    4	&	178.20	&	126.30	&	 309.55	 & \object{HD\,210752}  &	      17   &	      12   &	   229.60  &	   159.50  &	    697.15 \\
\object{HD\,113679} 	&	    6	&	    6	&	 92.60	&	 37.80	&	 192.95	 & \object{HD\,215257}  &	      37   &	      31   &	   288.30  &	   108.50  &	    976.52 \\
\object{HD\,11397 } 	&	   33	&	   33	&	151.40	&	 33.50	&	 649.63	 & \object{HD\,218504}  &	      15   &	      15   &	   192.00  &	    73.60  &	    582.42 \\
\object{HD\,119949} 	&	    5	&	    2	&	192.60	&	192.50	&	 272.31	 & \object{HD\,221580}  &	      54   &	      54   &	   127.20  &	    42.40  &	    649.58 \\
\object{HD\,121004} 	&	    5	&	    5	&	115.70	&	 50.00	&	 178.49	 & \object{HD\,223854}  &	       4   &	       3   &	   154.50  &	   134.30  &	    253.96 \\
\object{HD\,123517} 	&	    9	&	    6	&	 93.60	&	 70.30	&	 209.56	 & \object{HD\,224347}  &	       8   &	       5   &	   159.90  &	    33.50  &	    261.12 \\
\object{HD\,124785} 	&	   17	&	   17	&	131.20	&	 46.10	&	 397.18	 & \object{HD\,224817}  &	      30   &	      23   &	   190.70  &	   119.50  &	    726.06 \\
\object{HD\,126681} 	&	   14	&	   13	&	103.70	&	 49.50	&	 285.61	 & \object{HD\,22879}   &	      36   &	      19   &	   362.70  &	   224.30  &	   1280.67 \\
\object{HD\,126793} 	&	    7	&	    4	&	191.80	&	 94.30	&	 292.44	 & \object{HD\,25704}   &	      20   &	       8   &	   190.70  &	   124.60  &	    458.90 \\
\object{HD\,126803} 	&	    7	&	    6	&	 98.00	&	 43.20	&	 188.84	 & \object{HD\,31128}   &	      37   &	      37   &	   127.40  &	    38.70  &	    600.57 \\
\object{HD\,128340} 	&	    5	&	    3	&	140.30	&	 70.30	&	 177.67	 & \object{HD\,38510}   &	       5   &	       2   &	   151.90  &	   116.20  &	    191.25 \\
\object{HD\,128575} 	&	    2	&	    0	&	  0.00	&	  0.00	&	   0.00	 & \object{HD\,40865}   &	      30   &	      20   &	   169.70  &	   112.20  &	    599.71 \\
\object{HD\,129229} 	&	    5	&	    5	&	165.50	&	 47.10	&	 225.19	 & \object{HD\,51754}   &	      21   &	      21   &	   143.00  &	    50.00  &	    472.26 \\
\object{HD\,131653} 	&	    4	&	    4	&	111.20	&	 59.70	&	 186.37	 & \object{HD\,56274}   &	      14   &	      10   &	   222.60  &	   161.80  &	    611.40 \\
\object{HD\,134088} 	&	    4	&	    3	&	219.00	&	154.00	&	 311.59	 & \object{HD\,59984}   &	      45   &	      33   &	   498.40  &	   241.90  &	   1774.18 \\
\object{HD\,134113} 	&	   28	&	   23	&	184.30	&	 73.20	&	 611.32	 & \object{HD\,61902}   &	       7   &	       4   &	   193.50  &	   106.10  &	    270.13 \\
\object{HD\,134440} 	&	   10	&	   10	&	124.50	&	 52.60	&	 292.34	 & \object{HD\,62849}   &	      17   &	      17   &	   101.80  &	    49.50  &	    300.91 \\
\object{HD\,144589} 	&	   11	&	   11	&	 91.10	&	 45.80	&	 249.85	 & \object{HD\,68089}   &	       7   &	       7   &	    84.60  &	    44.80  &	    177.31 \\
\object{HD\,145344} 	&	    5	&	    4	&	139.20	&	 65.50	&	 187.81	 & \object{HD\,68284}   &	      10   &	       5   &	   220.40  &	   114.30  &	    371.61 \\
\object{HD\,145417} 	&	    5	&	    2	&	244.00	&	127.90	&	 275.49	 & \object{HD\,69611}   &	       6   &	       3   &	   188.10  &	   123.40  &	    260.23 \\
\object{HD\,147518} 	&	    4	&	    3	&	104.60	&	 60.20	&	 158.86	 & \object{HD\,75745}   &	      14   &	       9   &	   115.50  &	    76.60  &	    288.92 \\
\object{HD\,148211} 	&	   34	&	   30	&	248.10	&	 72.80	&	 826.44	 & \object{HD\,77110}   &	      16   &	      16   &	   143.20  &	    55.80  &	    458.88 \\
\object{HD\,148816} 	&	    7	&	    4	&	260.90	&	161.90	&	 413.57	 & \object{HD\,78747}   &	      26   &	      17   &	   237.80  &	   165.80  &	    832.53 \\
\object{HD\,149747} 	&	    9	&	    9	&	 93.00	&	 26.60	&	 166.58	 & \object{HD\,79601}   &	      16   &	      11   &	   215.10  &	   134.70  &	    588.14 \\
\object{HD\,150177} 	&	   30	&	   17	&	475.60	&	177.20	&	1135.11	 & \object{HD\,88474 }  &	       6   &	       3   &	   136.00  &	    63.70  &	    188.49 \\
\object{HD\,161265} 	&	    2	&	    1	&	 48.50	&	 48.50	&	  48.50	 & \object{HD\,88725}   &	      22   &	      16   &	   233.40  &	   185.50  &	    857.18 \\
\object{HD\,164500} 	&	    2	&	    2	&	 89.40	&	 81.60	&	 121.04	 & \object{HD\,90422}   &	       7   &	       4   &	   208.30  &	    85.60  &	    291.89 \\
\object{HD\,167300} 	&	    9	&	    9	&	109.70	&	 47.10	&	 255.80	 & \object{HD\,91345}   &	       8   &	       8   &	    94.80  &	    30.70  &	    160.61 \\
\object{HD\,16784 } 	&	    3	&	    1	&	160.60	&	160.60	&	 160.60	 & \object{HD\,94444}   &	       7   &	       4   &	   166.30  &	    90.00  &	    264.05 \\
\object{HD\,171028} 	&	   48	&	   39	&	184.80	&	 85.70	&	 835.80	 & \object{HD\,95860}   &	       7   &	       7   &	    94.40  &	    48.20  &	    180.14 \\
\object{HD\,171587} 	&	   14	&	   14	&	169.60	&	 75.60	&	 492.28	 & \object{HD\,967}     &	      34   &	      28   &	   175.90  &	   108.30  &	    770.54 \\
\object{HD\,175179} 	&	    3	&	    3	&	113.70	&	 91.00	&	 175.98	 & \object{HD\,97320 }  &	       6   &	       4   &	   183.50  &	    95.00  &	    263.43 \\
\object{HD\,17548 } 	&	   10	&	    9	&	185.90	&	 53.60	&	 252.61	 & \object{HD\,97783}   &	       6   &	       4   &	   132.00  &	   105.40  &	    227.50 \\
\object{HD\,175607} 	&	    7	&	    5	&	134.10	&	 85.90	&	 246.40	 & BD+062932	   &	       4   &	       4   &	    68.10  &	    51.00  &	    119.78 \\
\object{HD\,17865 } 	&	   21	&	   17	&	184.50	&	 98.60	&	 591.48	 & BD+063077	   &	       1   &	       1   &	    34.10  &	    34.10  &	     34.10 \\
\object{HD\,181720} 	&	   29	&	   21	&	228.20	&	114.10	&	 704.80	 & BD+083095	   &	       3   &	       3   &	    80.10  &	    44.00  &	    103.79 \\
\object{HD\,187151} 	&	    1	&	    1	&	105.00	&	105.00	&	 105.00	 & BD-004234	   &	       1   &	       1   &	    15.10  &	    15.10  &	     15.10 \\
\object{HD\,190984} 	&	   46	&	   44	&	158.40	&	 51.90	&	 718.18	 & BD-032525	   &	       2   &	       1   &	    52.10  &	    52.10  &	     52.10 \\
\object{HD\,193901} 	&	    3	&	    3	&	107.60	&	 57.20	&	 137.78	 & BD-084501	   &	       4   &	       4   &	    62.30  &	    37.10  &	    101.90 \\
\object{HD\,195633} 	&	    4	&	    2	&	160.40	&	 93.60	&	 185.71	 & CD-231087	   &	      35   &	      35   &	    84.50  &	    41.00  &	    421.16 \\
\object{HD\,196892} 	&	    3	&	    3	&	 98.80	&	 66.80	&	 150.62	 & CD-436810	   &	       9   &	       8   &	    88.20  &	    43.40  &	    171.55 \\
\object{HD\,197083} 	&	   12	&	   12	&	128.40	&	 52.30	&	 353.37	 & CD-4512460	   &	       4   &	       3   &	    52.10  &	    40.50  &	     81.31 \\
\object{HD\,197197} 	&	   21	&	   18	&	207.60	&	 83.20	&	 486.02	 & CD-452997	   &	       6   &	       1   &	    24.20  &	    24.20  &	     24.20 \\
\object{HD\,197536} 	&	    3	&	    2	&	157.80	&	129.00	&	 203.82	 & CD-571633	   &	       7   &	       5   &	   107.00  &	    66.00  &	    202.12 \\

\hline
\end{tabular}
\tablefoot{ \textit{totsp} is the total number of spectra observed; \textit{sumsp} is the number of spectra used for the combination of the final spectrum for each star; \textit{maxsn} is highest S/N value from the combined spectra for each star; \textit{minsn} is the lowest value from the combined spectra for each star ; and \textit{sumsn} is the final S/N for the combined spectrum.}
\end{scriptsize}
\label{tabspec}
\end{table*}

Several programs have been compiled to try to understand the distribution of the planets and the metallicity correlation. In particular, some 
are focused on metal-poor stars with the goal of not only checking the frequency of giants and low-mass planets in these stars, 
but also of measuring the lower limit in metallicity where it is possible to form and find giant planets. One of these programs is part of 
the HARPS GTO planet search program \citep[][]{Mayor-2003}. In this paper we present the precise derivation of fundamental spectroscopic stellar parameters 
and make an estimate of the masses for the stars in this sample.

We present a catalog of spectroscopic stellar parameters for the metal-poor sample observed with HARPS to search planets. In Sect. 2 we describe the observations with the 
HARPS spectrograph. Section 3 describes the procedure used to derive precise spectroscopic stellar parameters and to estimate for the stellar masses and 
new spectroscopic parallaxes based on the derived parameters. In Sect. 4, we compare our temperature values 
with the ones obtained with an IRFM (infra-red flux method) calibration to check for consistency. In Chapter 5 we redo a calibration for the temperature as a function of \textit{B-V} 
and [Fe/H] using the new parameters derived in this work that were added to data from a previous work. Finally in Chapter 6 we summarize the work presented here.

\section{The sample \& observations}

The sample of metal-poor stars, part of the HARPS GTO planet search program, is composed of a total of 104 stars. This sample was compiled from the 
catalog of \citet[][]{Nordstrom-2004}: late-F, G, and K stars (with b-y $>$ 0.33) with declination lower than $+10\,^{\circ}$ and 
visual V magnitude brighter than 12. All known visual and spectroscopic binaries were excluded, as were giants and stars that present 
a projected rotational velocity $v \sin i$ greater than 6 Km/s (to avoid active stars that normally rotate faster). The final selection of the targets was 
made by considering only the targets with photometric [Fe/H] between -0.5 and -1.5. We direct the reader to \citet[][]{Santos-2010b} for more details on the sample and a complete list of targets.

\begin{figure}[!t]
\centering
\includegraphics[width=8cm]{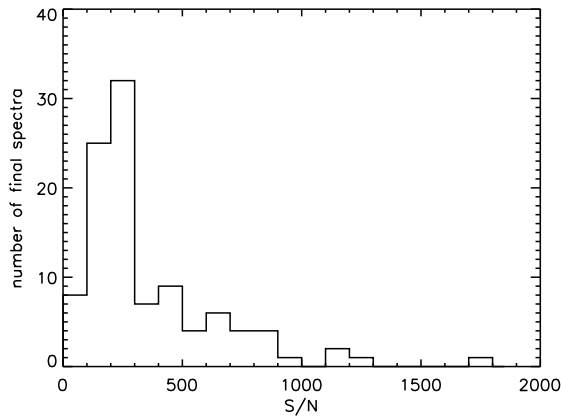}
\caption[]{Distribution of the signal-to-noise ratio of the final spectra compilation.}
\label{fig1}
\end{figure}

The data was collected between October 2003 and April 2010 with the HARPS spectrograph mounted in the ESO 3.6m telescope at La Silla, Chile. 
Since this sample was part of the HARPS GTO planet search program, most of the stars in this sample have several individual spectra collected through the duration of all the runs. 
The individual spectra of each star were reduced using the HARPS pipeline and then combined with IRAF\footnote{IRAF is distributed by National Optical Astronomy
Observatories, operated by the Association of Universities for Research in Astronomy, Inc., under contract with the National
Science Foundation, U.S.A.} after correcting for its radial velocity. The final spectra have very good quality with a resolution of R
$\sim$ 110 000 and S/N that vary from $\sim$ 30 to $\sim$ 2000, depending on the amount and quality of the original spectra. Figure \ref{fig1} shows the distribution 
of the final signal-to-noise ratio where $\sim$ 92 \% of the sample has S/N $>$ 100 and $\sim$ 68\% has S/N $>$ 200.

\section{Stellar parameters}

The spectroscopic stellar parameters and metallicities were derived following the same procedure as used in previous works \citep[][]{Santos-2004b,Sousa-2008}. 
The method is based on the equivalent widths of Fe I and Fe II weak lines, 
by imposing excitation and ionization equilibrium assuming LTE. For this task we used the 2002 version of 
the code MOOG \citep[][]{Sneden-1973} and a grid of Kurucz Atlas 9 plane-parallel model atmospheres \citep[][]{Kurucz-1993}. 
[Fe/H] is used as a proxy for the metallicity in this procedure.

The equivalent widths were automatically measured with the ARES\footnote{The ARES code can be downloaded at 
http://www.astro.up.pt/$\sim$sousasag/ares} code \citep[Automatic Routine for line Equivalent widths in stellar 
Spectra - ][]{Sousa-2007} which reproduces the common manual EWs measurements with success. The procedure used in \citet[][]{Sousa-2008} was closely followed, and the same 
input parameters were used for ARES in this work. Since in this sample there is a significant number of stars with S/N lower than 100 and even a couple of spectra with S/N 
lower than 50, we present an empirical formula that upgrades Table 2 presented in \citet[][]{Sousa-2008} for low S/N levels of the spectra. This formula can be used to 
obtain the recommended values for the ARES parameter \textit{rejt} for low S/N values. These values can be obtained with the empirical equation
\begin{equation}
rejt = 0.948378 + 6.39270e^{-4}(S/N)-2.41632e^{-6}(S/N)^2
\end{equation}

This empirical relation between the S/N and the \textit{rejt} parameter was obtained by considering a subsample of 6 HARPS 
spectra selected from the metal poor sample studied in this work. The spectra were selected in a way to represent different low S/N values, ranging between 30 and 100, and then setting 
for each spectra, by eye, the best value of \textit{rejt} that would fit correctly the local continuum. To verify this task we selected a few isolated lines and set the ARES code to 
allow the presentation of plots for the local automatic normalization and also automatic determination of the continuum.
These subjective values were then used to adjust a simple polynomial of second order. With this equation and the previous recommended values for the \textit{rejt} parameter 
for S/N over 100, we are ready to measure automatically and in a systematic way the equivalent widths for the several absorption lines in the stellar spectra. These values and this new 
equation are valid for HARPS spectra; however, these values can be used as a first approach and a similar procedure should be made to obtain compatible values to be used for other resolution spectrographs.

The spectroscopic parameters are presented in Table \ref{tab2}. We compared these values with the ones we could find in the work of \citet[][]{Fischer_Valenti-2005}. 
From the 9 stars we found in common we observe a mean difference of 
$-95 \pm 90$ K, $-0.03 \pm 0.15$ dex, and $-0.06 \pm 0.05$ for temperature, surface gravity, and [Fe/H], respectively.

\subsection{Precision errors vs. acuracy errors}

Table \ref{tab2} also presents the errors determined following the same procedure as described in previous works \citep[][]{Sousa-2008, Santos-2004b}. These values are 
indeed precision errors that are intrinsic to the spectroscopic method used in this work. These values are typically very small, especially for the stars more like the Sun. This comes 
directly from the method itself since a differential analysis is performed with the Sun as a reference. Stars that are significantly cooler or hotter than the Sun will have larger intrinsic errors. 
If the reader is interested in accurate errors, then we also estimate possible systematic errors. These systematic errors can be estimated when comparing the parameters derived using a given method with others 
derived from different methods. Using the comparison plots in \citet[][]{Sousa-2008}, we can assume a systematic error of 60 K coming from the comparison between our method temperatures and the ones derived with the 
IRFM (Figure 3, (g) and (h)). For the surface gravity we can consider a systematic error of 0.1 dex coming from an average of the comparison of the different methods (Figure 4). Similarly, an average of 0.04 dex 
extracted from the different methods in Figure 5 of the same work can be used for the systematic error for [Fe/H].

These values can be quadratically added to the precision errors. For the stars below 5000 K the systematic values should be higher. As discussed before, these stars are farthest 
from the Sun, so any systematic will be more significant. As an example, for stars below 5000 K a more appropriate typical systematic error will be closer to 100 K for the temperature.

\begin{figure}[!t]
\centering
\includegraphics[width=8cm]{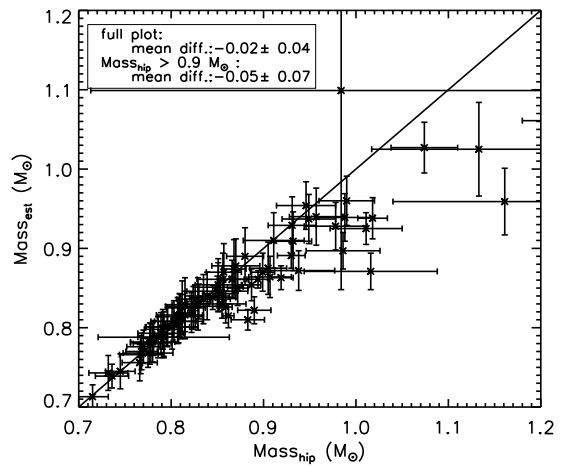}
\caption[]{Comparison between the estimated masses using the Hipparcos parallaxes and the estimated masses using the iterative procedure. 
the mean diference and dispersion are also shown for stars with mass greater than 0.9 solar masses.}
\label{figmass}
\end{figure}

\subsection{Masses and parallaxes}

Stellar masses were estimated as in previous works \citep[e.g.][]{Santos-2004b, Sousa-2008}. In this case we applied the stellar evolutionary models from the Padova group, 
computed using the web interface dealing with stellar isochrones and their derivatives (http://stev.oapd.inaf.it/cgi-bin/cmd) to the stars of our sample. For this task we used 
the Hipparcos parallaxes and V magnitudes \citep[][]{Hipparcus-2007}, a bolometric correction from \citet[][]{Flower-1996}, and the effective temperature derived from the spectroscopic analysis. 
The errors presented for the masses were also given by the web interface.

Since the parallaxes for most of these stars present large erros ($\sim$ 10-20\%) and for some cases they even present absolute errors of the same 
order of magnitude as the parallaxes, we used an iterative process to derive new values for the parallaxes and therefore also for the masses. 
This iterative process was already used in \citet[][]{Santos-2010} where it was applied for a single star.

The iterative procedure in this work makes use of Eq. (1) in \citet[][]{Santos-2004b}, the relation between luminosity, radius, and parallax, and the Padova web-interface to derive the masses. 
\begin{enumerate}
\item  First, we fixed the bolometric correction and the visual magnitude of the star to the values derived by the calibration of Flower (1996) and 
the value listed in the Hipparcos catalog, respectively. 
An initial value for the stellar mass was also obtained using the Hipparcos parallax, the V magnitude, the metallicity, and the temperature derived 
from spectroscopy. However, since in our sample there are some stars without measured Hipparcos parallaxes (\object{HD\,62849}, \object{HD\,75745}, \object{HD\,95860}, \object{HD\,123517}, 
\object{HD\,144589}, \object{HD\,149747}, \object{HD\,171028}, \object{CD-4512460}, and \object{CD-452997}), it was not possible to estimate the mass 
to initiate the first step of the procedure. To overcome this problem we alternatively used the calibration presented in the work of 
\citet[][]{Torres-2010} to derive the initial guess for the mass for each star and then allowed initiation of iterative process with this value instead.
\item The second step is to obtain a new value for the parallax that is used in the next iteration to derive a new mass using the web interface from Padova.
\end{enumerate}
This procedure is then followed iteratively until we find a convergence for both the mass and parallax. On average, for all the stars we found a 
convergence after only three iterations. 

In Fig. \ref{figmass} a comparison is presented between the masses derived using the Hipparcos parallaxes and the mass estimated using the iterative procedure. 
From this figure we can see that both mass values are compatible within the errors with a increased dispersion for higher values where the errors also 
increase significantly.
The final value for the parallax can be significantly different from the one measured by Hipparcos.
There were a few stars (4) for which it was not possible to derive any value for the mass and/or parallax since the web interface would 
not return any values. The reason for this is unknown to us.

\begin{figure}[!t]
\centering
\includegraphics[width=8cm]{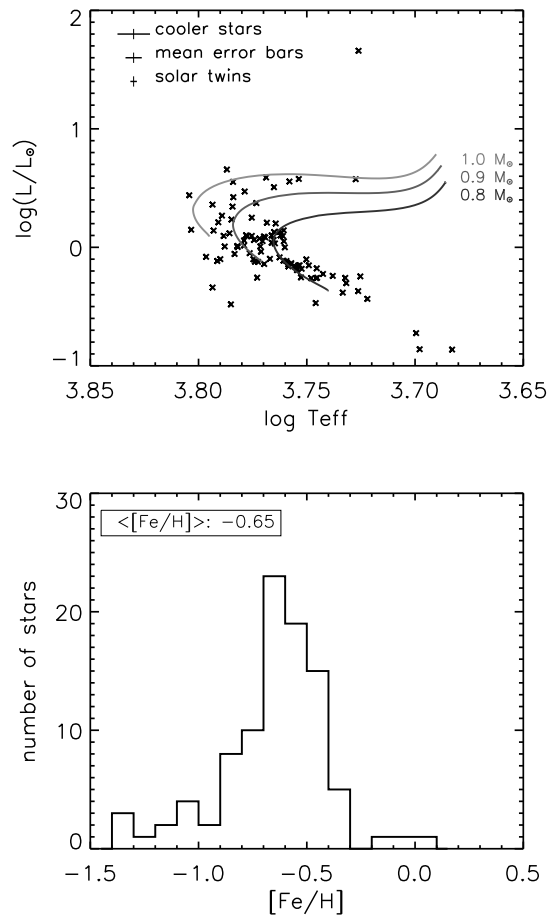}
\caption[]{In the top panel, we present the distribution of the sample stars in the H-R diagram. We also plot some evolutionary tracks computed with CESAM for a 0.8, 0.9 and 
1.0 M$_{\sun}$ assuming the mixing length parameter as 1.4, the initial Helium ratio Y = 0.26 and for [Fe/H] $\sim$ -0.5. 
In the bottom panel, we present the metallicity distribution of this sample.}
\label{fighr}
\end{figure}

The luminosity was computed by considering the estimated Hipparcos parallaxes, V magnitude and the bolometric correction. Its error is derived from the parallax errors that are the main source 
of uncertainty in calculating luminosity. The typical error for the luminosity is $\sim 0.04$, which is obtained by assuming the mean parallax for the stars ($\sim$20 mas) and a typical error 
for the estimated parallaxes of 1 mas.

Figure \ref{fighr} presents some characteristics of the sample. The top plot shows the distribution of the sample on the Hertzsprung-Russell diagram, where we represent evolutionary tracks for
0.8, 0.9, and 1.0 M$_{\sun}$, computed with the CESAM code \citep[][]{Morel-1997} using the mixing length parameter with the value 1.4, the initial helium content as Y = 0.26 and assuming a Z that 
corresponds to the mean observed metallicity for the stars corresponding to [Fe/H] $\sim$ -0.5. For this plot we made use of the estimated paralaxes because we have these values for a higher number of stars in the sample. 
The typical error boxes are also presented in this specific diagram, where the error in the luminosity is the same as described before, and the error in the temperature is derived 
from the spectroscopic method. This plot shows that the sample is composed mainly of main-sequence solar-type stars. We can see a dispersion in this figure that can be explained by the uncertainty of 
the parallaxes that are very small and can lead to larger errors on the luminosity. 
In the bottom plot we present the metallicity distribution that has a mean value of about -0.65 dex. This shows that these stars belong to an excellent sample that is perfectly complementary 
in terms of metallicity to the other samples used to search for planets. As a reference, the mean metallicity of the stars in the main HARPS GTO planet search 
program is -0.09 \citep[][]{Sousa-2008}. 
This sample is ideal for probing and testing the connection between metal-poor stars and the existence of planets around them.

In Table \ref{tab2} we present the derived spectroscopic parameters for 97 stars, together with the determined parallaxes and masses. The absent stars are explained in the next section.

\subsection{Special cases}

There are 7 stars for which we were unable to derive spectroscopic stellar parameters. HD161265, BD-032525, HD164500, and HD187151 were excluded because they 
are double-line spectroscopic binaries SB2 stars. BD-004234 is suspected of also being a spectroscopic binary, so, it was also removed from the observation list from the runs. 
HD197890 was observed a few times but presents a strange spectrum. This could possibly be an active star. 
HD128575 was observed 2 times, also indicating strange spectral features, possibly an active star. Both these stars do not allow combining the individual spectra, so 
the parameters were not derived. A complete description of these special cases can be found in Santos et al. (2010).

HD967 is a target also present in the main HARPS GTO program with the parameters derived in \citet[][]{Sousa-2008}. The parameters presented before 
(T$_{\mathrm{eff}}$: 5568 K ; $\log{g}$: 4.51; [Fe/H]: -0.68)  are extremely consistent with the ones presented in this work, proving that the spectrocopic analysis performed 
by our team is precise and systematic.

\subsection{High surface gravity}

In Table \ref{tab2} we observe that there are a few stars (6) with high surface gravity values (log g $>$4.7). 
These kinds of values are not expected, and stellar interior models may encounter dificulties when trying to fit this kind of surface gravity for dwarf stars.
A more attentive reader can find out that these high values are more or less correlated with the low metallicity for these stars. In fact, from 
these stars only BD+063077 is above [Fe/H] $\sim$ -1. For this specific star we may explain that the high value of gravity comes from the very 
low S/N of the spectrum.

For the rest of these stars, they all have $[Fe/H] < -1$. We do not rule out that there might be some 
systematic result of our spectroscopic method. For these stars the lines are all typically very weak, some even 
undetected due to the low amount of iron. Therefore for these stars we have on average fewer lines used for determining the stellar 
parameters. Moreover, since the lines are typically weaker, they propagate higher errors coming from the equivalent width measurements. For these stars 
we compared the stellar parameters directly with others in the literature. As an example, for 4 of these 6 stars, that we found in common in the work of 
\citet[][]{Casagrande-2010}, we see a mean difference of $-28 \pm 76$ K for temperature. Although there is a large dispersion for all the different methods, the values are 
mostly consistent and within the errors. Even for surface gravity we typically found high values for these stars in the other methods. Although considering that the gravity may be 
overestimated using our method for these low metal stars, the other parameters seems to be reliable. A more complete comparison is made in the next section, which shows 
that there is no clear metallicity correlation in the comparison. Nevertheless the reader should take possible systematics into account for these high surface gravity stars.

\begin{figure}[!t]
\centering
\includegraphics[width=8cm]{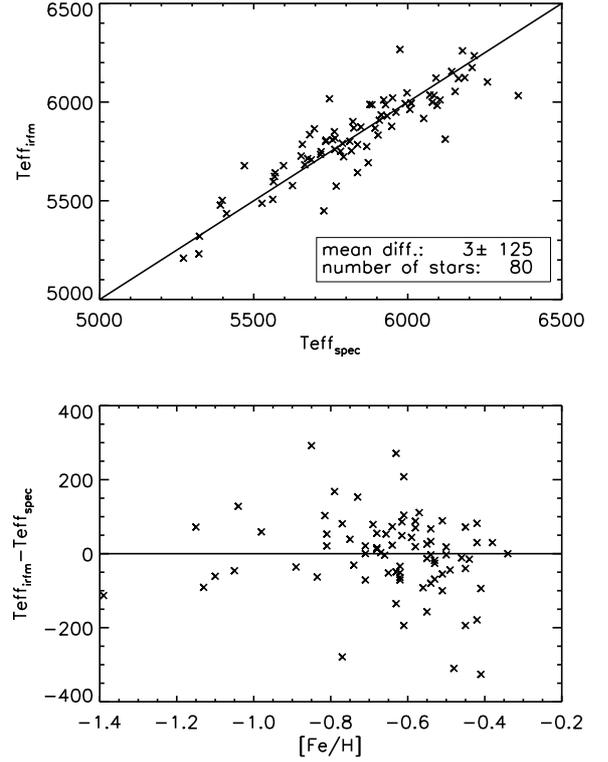}
\caption[]{Comparison between the derived spectroscopic temperatures and the ones derived using an IRFM calibration.}
\label{figirfm}
\end{figure}

\section{IRFM calibration comparison}

To check the consistency of the derived parameters we compared the derived spectroscopic temperatures 
and the ones derived using an IRFM calibration presented in the work of \citet[][]{Casagrande-2010}. 
To perform this comparison we first need to obtain the values of the several photometric colors required for the IRFM calibration. 
These colors were compiled from different sources. The U was obtained from the SKY2000 Catalog, Version
 4 \citep[][]{Myers-2001}, and the B, V, R were obtained from the NOMAD catalog \citep[][]{Zacharias-2004}. 
The I was calculated from the index V-I obtained in the Hipparcos catalog \citep[][]{Hipparcus-2007}. Finally the JHK colors were obtained from the 
2MASS catalog \citep[][]{2MASS-2003}. The list of the colors used in this work are presented in Table \ref{tabphoto}.

For each of the stars, and considering the available colors, we derived several values for the absolute flux calibration and the angular diameter 
calibrations from the coefficients presented in both Tables 5 and 6 from the work of \citet[][]{Casagrande-2010}. These values were then average-weighted 
considering the errors in the tables for each color calibration to obtain both $f_{bolmean}$ and $\theta_{radmean}$. With this it is possible to compute the calibrated 
effective temperature ($T_{\mathrm{eff\_irfm}}$) using the equation
\begin{equation}
T_{\mathrm{eff\_irfm}} = \sqrt{ \frac{ 2 \sqrt{f_{bolmean}/\sigma} }{\theta_{radmean}} }   
\end{equation}
where $\sigma$ is the Stefan$–$Boltzmann constant.

Figure \ref{figirfm} shows the comparison between the derived spectroscopic temperatures and the temperatures obtained with this IRFM calibration. The comparison is consistent when presenting a mean difference 
of 3 $\pm$ 125 K. This shows that both the spectroscopic temperatures and the temperatures derived from this IRFM are on the same scale of temperature.

In the bottom plot of Figure \ref{figirfm} we present the difference between the derived effective temperatures as a function of metallicity. We cannot see any clear metallicity effect in this plot, meaning that any 
NLTE effect for low metallicity is not significant for deriving of the temperature. The high dispersion may be explained by the possibly bad quality of the photometry used for the IRFM calibration.

\begin{figure}[!t]
\centering
\includegraphics[width=8cm]{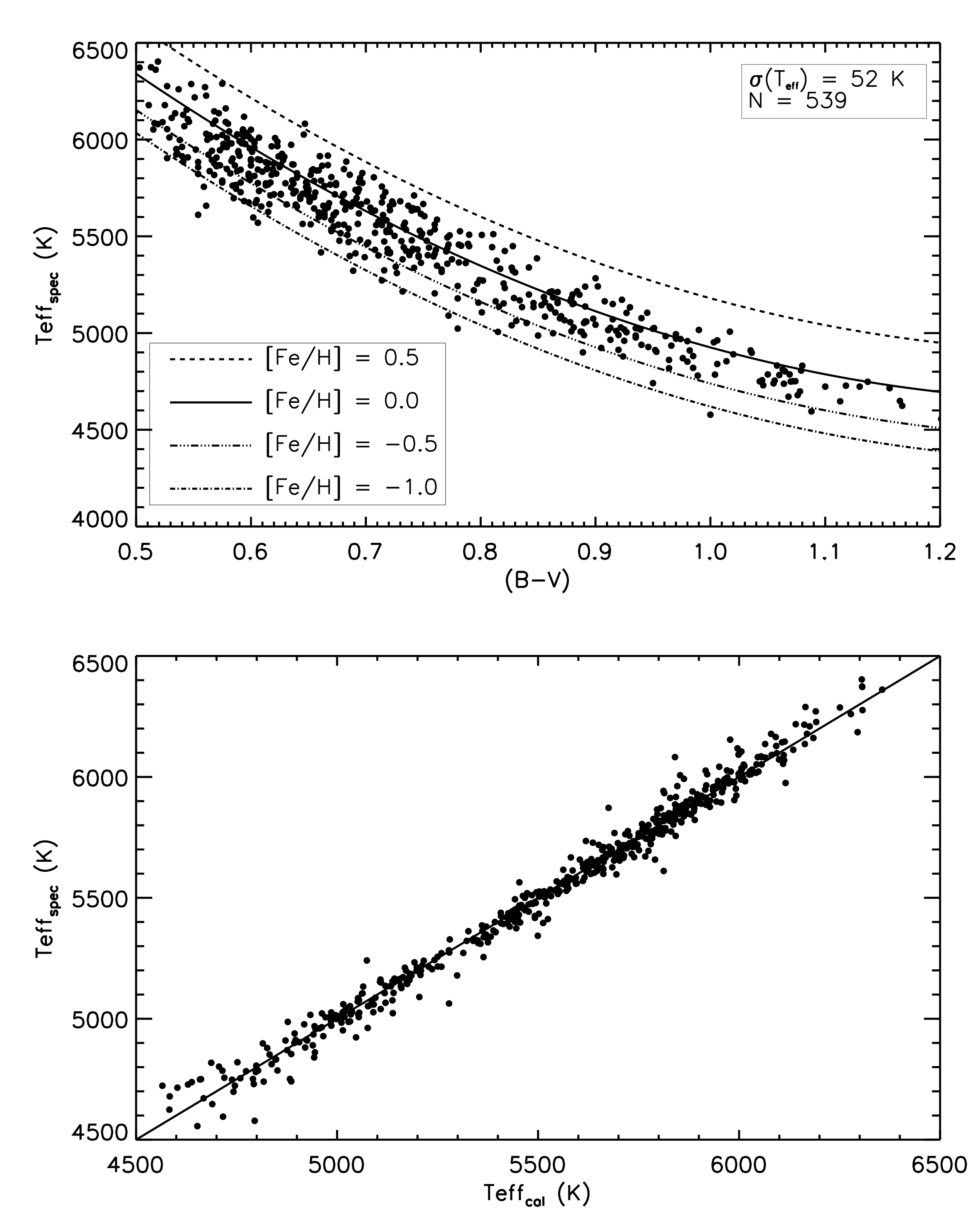}
\caption[]{Calibration of the effective temperature as a function of the color index \textit{B-V} and [Fe/H]. The 4 fitted 
lines correspond to lines with constant values of [Fe/H] (-1.0, -0.5, 0.0, 0.5). The bottom panel presents the 
direct comparison between the spectroscopic temperatures and the calibration. The filled line is the identity line. }
\label{fig_teff_bv_feh}
\end{figure}

\section{A new calibration for effective temperature}

In this section we present a new calibration for the effective temperature as a function of \textit{B-V} and [Fe/H]. This kind of calibration was already presented in \citet[][]{Sousa-2008}. 
Here we derive a new calibration by adding the data derived for the metal-poor sample. This will increase not only the number of stars but, more importantly, widen the metallicity 
range for the calibration. Therefore by adding these new metal-poor stars we achieve a new metallicity interval that ranges from the values -1.5 to 0.5. 

This new calibration made use of the spectroscopic parameters derived in both works, and used the $B-V$ value taken from the Hipparcos catalog \citep[][]{ESA-1997}.
The result is illustrated in Fig. \ref{fig_teff_bv_feh}, and the calibration is then expressed by
\begin{equation}
\label{eqcal}
\begin{split}
  T_{\mathrm{eff}} = 8939 &- 6395(B{-}V) + 2381(B{-}V)^2 +\\
& + 451[Fe/H] + 154[Fe/H]^2.
\end{split}
\end{equation}

The final coefficients were obtained after removing an outlier, i.e., after the first fit to all the stars, we removed the stars that presented a difference greater than 4 sigma of the fit. There were 6 outliers, HD62849, 
HD88474, HD119949, HD128340, BD-084501, and CD-452997. The reasons for these outliers maybe errors in the the color index \textit{B-V} and/or the errors in the derived effective 
temperature, especially for the cases of BD-084501 (lower number of analyzed lines and one of the most metal-poor stars in the sample) and CD-452997 ( S/N of the spectra is $\sim$ 25).


The standard deviation for the fit is only 52 K, illustrating the good quality of the relation. This result is similar to the previous calibration presented
 in \citet[][]{Sousa-2008}.
and the calibration can be useful and be applied to stars without the need for a detailed spectroscopic analysis, with the guarantee that the result will lie on the same effective temperature scale. 
This calibration is valid in the following intervals: $4500 K < T_{eff} <6400 K$, $-1.5 < [Fe/H] < 0.50$, and $ 0.40 < B{-}V < 1.20$.


\section{Conclusions}

In this work we presented precise stellar parameters for a sample of metal-poor stars. The stellar parameters were derived in a consistent way following 
the same method as in previous works. It is crucial that the spectroscopic parameters for these stars are derived precisely and in a systematic way to allow 
correct comparison between stars. This is very useful when searching for clues for the stellar and planet formations that are typically done by comparing the stars with detected planets 
to the single stars that do not show any evidence of hosting any planet. These parameters and abundances will be used to study the frequency of planets as a function of the stellar parameters, but this is
beyond the scope of this paper.

ARES is a very important tool for this task, not only because it is an automatic tool that allows faster and more completely analysis of the spectra of many stars, but more importantly, because it clearly allows 
eliminating of most of the human factor that was creating larger errors in the spectral analysis (more specifically the subjective position of the continuum located by eye when measuring the EWs of 
the lines with interactive routines). 

We also present estimations for the mass of these stars using similar procedures as in previous works. Here it was necessary to overtake the problem of the high errors in the Hipparcos parallaxes 
presented for most of the stars in the sample. Therefore we estimated a second value for the mass by assuming a different parallax based on the derived spectroscopic parameters.

The effective temperature for these metal-poor stars was tested and compared against an IRFM calibration. The comparison between the two different approaches to derive the effective temperature are consistent, 
meaning that our spectroscopic method is still valid for lower metallicity stars.

Finally a new calibration for the effective temperature as a function of the color index \textit{B-V} and [Fe/H] is presented where the metallicity range is now wider thanks to using 
the parameters derived in this work.

\begin{acknowledgements}
S.G.S acknowledges the support from the Funda\c{c}\~ao para a Ci\^encia e Tecnologia (Portugal) in the form of grant SFRH/BPD/47611/2008. NCS thanks for the 
support by the European Research Council/European Community under the FP7 through a Starting Grant, as well as the support from the Funda\c{c}\~ao para a Ci\^encia 
e a Tecnologia (FCT), Portugal, through program Ci\^encia\,2007. We also acknowledge support from the FCT in the form of grants reference PTDC/CTE-AST/098528/2008, 
PTDC/CTE-AST/66181/2006, and PTDC/CTE-AST/098604/2008.

\end{acknowledgements}

\bibliographystyle{aa}
\bibliography{15646}

\longtab{2}{
\begin{scriptsize}
\begin{longtable}{cccccccccc}
\caption{\label{tab2} Stellar parameters and respective errors derived for the metal poor sample.}\\
\hline\hline
Name & T$_{\mathrm{eff}}$ (K)& $\log{g}_{spec}$ & $\xi_{\mathrm{t}}$ (km/s) & [Fe/H] &  N(\ion{Fe}{i},\ion{Fe}{ii}) & $Mass_{hip}$ ($M_{\sun}$) & $Mass_{est}$ ($M_{\sun}$) & $\pi_{phip}$ & $\pi_{pest}$\\
\hline
\endfirsthead
\caption{continued.}\\
\hline\hline
Name & T$_{\mathrm{eff}}$ (K)& $\log{g}_{spec}$ & $\xi_{\mathrm{t}}$ (km/s) & [Fe/H] &  N(\ion{Fe}{i},\ion{Fe}{ii}) & $Mass_{hip}$ ($M_{\sun}$) & $Mass_{est}$ ($M_{\sun}$) & $\pi_{phip}$ & $\pi_{pest}$\\
\hline
\endhead
\hline
\endfoot
\object{HD\,967}       	&	5568	$\pm$	    17	&	4.53	$\pm$	  0.02	&	0.77	$\pm$	  0.04	&	-0.68 $\pm$   0.01   &        248,33	&	0.78   $\pm$   0.02   &       0.78   $\pm$   0.03   &	    23.50   $\pm$   1.02    &	    26.77\\
\object{HD\,11397}     	&	5564	$\pm$	    26	&	4.46	$\pm$	  0.04	&	0.75	$\pm$	  0.05	&	-0.54 $\pm$   0.02   &        256,33	&	0.80   $\pm$   0.03   &       0.81   $\pm$   0.03   &	    19.03   $\pm$   1.36    &	    18.46\\
\object{HD\,16784}     	&	5837	$\pm$	    22	&	4.34	$\pm$	  0.02	&	1.14	$\pm$	  0.04	&	-0.65 $\pm$   0.02   &        240,34	&	0.86   $\pm$   0.03   &       0.83   $\pm$   0.02   &	    17.71   $\pm$   0.92    &	    21.76\\
\object{HD\,17548}     	&	6011	$\pm$	    26	&	4.44	$\pm$	  0.02	&	1.18	$\pm$	  0.04	&	-0.53 $\pm$   0.02   &        238,33	&	0.88   $\pm$   0.02   &       0.89   $\pm$   0.04   &	    18.19   $\pm$   0.72    &	    20.59\\
\object{HD\,17865}     	&	5877	$\pm$	    24	&	4.32	$\pm$	  0.03	&	1.16	$\pm$	  0.04	&	-0.57 $\pm$   0.02   &        242,33	&	0.89   $\pm$   0.01   &       0.85   $\pm$   0.02   &	    15.60   $\pm$   0.65    &	    19.22\\
\object{HD\,22879}     	&	5884	$\pm$	    33	&	4.52	$\pm$	  0.03	&	1.20	$\pm$	  0.07	&	-0.82 $\pm$   0.02   &        222,34	&	0.81   $\pm$   0.01   &       0.81   $\pm$   0.03   &	    39.12   $\pm$   0.56    &	    49.18\\
\object{HD\,25704}     	&	5942	$\pm$	    33	&	4.52	$\pm$	  0.02	&	1.37	$\pm$	  0.07	&	-0.83 $\pm$   0.02   &        218,34	&	0.82   $\pm$   0.01   &       0.82   $\pm$   0.03   &	    19.43   $\pm$   0.66    &	    24.75\\
\object{HD\,31128}     	&	6096	$\pm$	    67	&	4.90	$\pm$	  0.05	&	3.02	$\pm$	  0.78	&	-1.39 $\pm$   0.04   &        124,24	&	0.78   $\pm$   0.03   &       0.77   $\pm$   0.02   &	    15.00   $\pm$   1.13    &	    23.23\\
\object{HD\,38510}     	&	5914	$\pm$	    37	&	4.32	$\pm$	  0.03	&	1.30	$\pm$	  0.07	&	-0.81 $\pm$   0.02   &        223,34	&	0.86   $\pm$   0.01   &       0.82   $\pm$   0.02   &	    15.49   $\pm$   0.78    &	    19.03\\
\object{HD\,40865}     	&	5719	$\pm$	    16	&	4.50	$\pm$	  0.03	&	0.87	$\pm$	  0.03	&	-0.38 $\pm$   0.01   &        255,33	&	0.86   $\pm$   0.03   &       0.86   $\pm$   0.03   &	    19.64   $\pm$   0.72    &	    20.52\\
\object{HD\,51754}     	&	5848	$\pm$	    24	&	4.49	$\pm$	  0.02	&	1.05	$\pm$	  0.04	&	-0.55 $\pm$   0.02   &        250,33	&	0.85   $\pm$   0.03   &       0.85   $\pm$   0.03   &	    12.56   $\pm$   1.16    &	    15.97\\
\object{HD\,56274}     	&	5734	$\pm$	    22	&	4.51	$\pm$	  0.03	&	0.94	$\pm$	  0.04	&	-0.54 $\pm$   0.02   &        252,33	&	0.84   $\pm$   0.03   &       0.84   $\pm$   0.03   &	    30.95   $\pm$   0.75    &	    31.26\\
\object{HD\,59984}     	&	5962	$\pm$	    27	&	4.18	$\pm$	  0.02	&	1.45	$\pm$	  0.05	&	-0.69 $\pm$   0.02   &        225,33	&	0.89   $\pm$   0.01   &       0.86   $\pm$   0.01   &	    35.82   $\pm$   0.54    &	    44.66\\
\object{HD\,61902}     	&	6209	$\pm$	    30	&	4.38	$\pm$	  0.03	&	1.58	$\pm$	  0.06	&	-0.62 $\pm$   0.02   &        212,33	&	0.93   $\pm$   0.02   &       0.91   $\pm$   0.04   &	    12.71   $\pm$   0.51    &	    17.07\\
\object{HD\,62849}     	&	5338	$\pm$	    20	&	3.59	$\pm$	  0.03	&	1.04	$\pm$	  0.02	&	-0.17 $\pm$   0.02   &        262,35	&	       -	      &       1.24   $\pm$   0.14   &		    -		    &	     5.47\\
\object{HD\,68089}     	&	5597	$\pm$	    27	&	4.53	$\pm$	  0.04	&	0.66	$\pm$	  0.07	&	-0.77 $\pm$   0.02   &        248,34	&	0.77   $\pm$   0.02   &       0.77   $\pm$   0.02   &	    14.17   $\pm$   0.86    &	    15.08\\
\object{HD\,68284}     	&	5933	$\pm$	    26	&	4.08	$\pm$	  0.03	&	1.40	$\pm$	  0.04	&	-0.50 $\pm$   0.02   &        245,33	&	1.01   $\pm$   0.04   &       0.93   $\pm$   0.02   &	    13.14   $\pm$   0.88    &	    16.57\\
\object{HD\,69611}     	&	5762	$\pm$	    25	&	4.31	$\pm$	  0.03	&	0.99	$\pm$	  0.04	&	-0.58 $\pm$   0.02   &        251,33	&	0.85   $\pm$   0.03   &       0.84   $\pm$   0.02   &	    20.50   $\pm$   0.95    &	    24.43\\
\object{HD\,75745}     	&	5885	$\pm$	    35	&	4.29	$\pm$	  0.03	&	1.34	$\pm$	  0.06	&	-0.78 $\pm$   0.03   &        226,34	&	       -	      &       0.81   $\pm$   0.02   &		    -		    &	    10.54\\
\object{HD\,77110}     	&	5717	$\pm$	    20	&	4.48	$\pm$	  0.02	&	0.86	$\pm$	  0.04	&	-0.50 $\pm$   0.02   &        253,33	&	0.84   $\pm$   0.02   &       0.84   $\pm$   0.03   &	    16.28   $\pm$   0.94    &	    18.28\\
\object{HD\,78747}     	&	5788	$\pm$	    20	&	4.44	$\pm$	  0.02	&	1.10	$\pm$	  0.04	&	-0.67 $\pm$   0.02   &        238,32	&	0.81   $\pm$   0.00   &       0.82   $\pm$   0.03   &	    24.53   $\pm$   0.56    &	    28.75\\
\object{HD\,79601}     	&	5825	$\pm$	    25	&	4.32	$\pm$	  0.03	&	1.09	$\pm$	  0.04	&	-0.59 $\pm$   0.02   &        247,34	&	0.85   $\pm$   0.02   &       0.84   $\pm$   0.02   &	    17.70   $\pm$   0.64    &	    21.19\\
\object{HD\,88474}     	&	6122	$\pm$	    40	&	3.91	$\pm$	  0.03	&	1.91	$\pm$	  0.07	&	-0.48 $\pm$   0.03   &        234,34	&	1.23   $\pm$   0.05   &       1.06   $\pm$   0.06   &	    6.51    $\pm$   0.53    &	     8.44\\
\object{HD\,88725}     	&	5654	$\pm$	    17	&	4.49	$\pm$	  0.03	&	0.86	$\pm$	  0.03	&	-0.64 $\pm$   0.01   &        245,33	&	0.80   $\pm$   0.01   &       0.80   $\pm$   0.03   &	    28.24   $\pm$   0.72    &	    32.10\\
\object{HD\,90422}     	&	6085	$\pm$	    33	&	4.14	$\pm$	  0.03	&	1.67	$\pm$	  0.06	&	-0.62 $\pm$   0.02   &        221,33	&	0.99   $\pm$   0.04   &       0.90   $\pm$   0.02   &	    10.15   $\pm$   0.84    &	    13.43\\   
\object{HD\,91345}     	&	5658	$\pm$	    39	&	4.53	$\pm$	  0.04	&	0.71	$\pm$	  0.12	&	-1.04 $\pm$   0.03   &        225,31	&	0.75   $\pm$   0.02   &       0.75   $\pm$   0.02   &	    16.71   $\pm$   0.89    &	    19.24\\   
\object{HD\,94444}     	&	5998	$\pm$	    27	&	4.34	$\pm$	  0.03	&	1.29	$\pm$	  0.05	&	-0.62 $\pm$   0.02   &        235,34	&	0.87   $\pm$   0.01   &       0.86   $\pm$   0.02   &	    17.38   $\pm$   0.77    &	    19.28\\   
\object{HD\,95860}     	&	6054	$\pm$	    25	&	4.48	$\pm$	  0.03	&	1.25	$\pm$	  0.04	&	-0.31 $\pm$   0.02   &        254,35	&	       -	      &       0.96   $\pm$   0.04   &		    -		    &	     9.96\\   
\object{HD\,97320}     	&	6165	$\pm$	    52	&	4.57	$\pm$	  0.04	&	1.50	$\pm$	  0.17	&	-1.05 $\pm$   0.03   &        183,31	&	0.82   $\pm$   0.01   &       0.83   $\pm$   0.03   &	    18.36   $\pm$   0.58    &	    23.31\\   
\object{HD\,97783}     	&	5682	$\pm$	    24	&	4.50	$\pm$	  0.02	&	0.88	$\pm$	  0.04	&	-0.73 $\pm$   0.02   &        243,33	&	0.79   $\pm$   0.02   &       0.79   $\pm$   0.03   &	    15.02   $\pm$   1.09    &	    17.80\\   
\object{HD\,102200}    	&	6185	$\pm$	    65	&	4.59	$\pm$	  0.04	&	1.52	$\pm$	  0.23	&	-1.10 $\pm$   0.04   &        167,30	&	0.81   $\pm$   0.02   &       0.83   $\pm$   0.03   &	    13.00   $\pm$   0.98    &	    18.08\\   
\object{HD\,104800}    	&	5697	$\pm$	    25	&	4.47	$\pm$	  0.02	&	0.87	$\pm$	  0.05	&	-0.79 $\pm$   0.02   &        235,34	&	0.78   $\pm$   0.02   &       0.78   $\pm$   0.02   &	    14.24   $\pm$   1.33    &	    15.93\\   
\object{HD\,105004}    	&	5756	$\pm$	    39	&	4.33	$\pm$	  0.03	&	0.80	$\pm$	  0.08	&	-0.81 $\pm$   0.03   &        232,33	&	0.79   $\pm$   0.07   &       0.79   $\pm$   0.03   &	    0.23    $\pm$   2.91    &	     8.39\\   
\object{HD\,107094}    	&	5562	$\pm$	    17	&	4.54	$\pm$	  0.03	&	0.74	$\pm$	  0.03	&	-0.51 $\pm$   0.01   &        257,33	&	0.81   $\pm$   0.03   &       0.81   $\pm$   0.03   &	    17.85   $\pm$   1.16    &	    18.73\\   
\object{HD\,108564}    	&	4818	$\pm$	    69	&	4.67	$\pm$	  0.17	&	0.26	$\pm$	  0.64	&	-0.97 $\pm$   0.07   &        210,13	&	0.70   $\pm$   0.00   & 	     -  	    &	    36.78   $\pm$   1.01    &	    -	 \\   
\object{HD\,109310}    	&	5922	$\pm$	    19	&	4.55	$\pm$	  0.02	&	1.15	$\pm$	  0.03	&	-0.51 $\pm$   0.01   &        245,34	&	0.87   $\pm$   0.03   &       0.88   $\pm$   0.03   &	    18.44   $\pm$   0.91    &	    22.17\\   
\object{HD\,109684}    	&	5992	$\pm$	    18	&	4.38	$\pm$	  0.02	&	1.22	$\pm$	  0.03	&	-0.34 $\pm$   0.01   &        254,35	&	0.93   $\pm$   0.03   &       0.93   $\pm$   0.04   &	    12.90   $\pm$   1.00    &	    14.67\\   
\object{HD\,111515}    	&	5398	$\pm$	    18	&	4.47	$\pm$	  0.02	&	0.71	$\pm$	  0.04	&	-0.61 $\pm$   0.01   &        256,32	&	0.77   $\pm$   0.02   &       0.78   $\pm$   0.02   &	    30.71   $\pm$   0.74    &	    30.27\\   
\object{HD\,111777}    	&	5666	$\pm$	    19	&	4.46	$\pm$	  0.03	&	0.82	$\pm$	  0.04	&	-0.68 $\pm$   0.01   &        249,33	&	0.80   $\pm$   0.02   &       0.80   $\pm$   0.03   &	    20.67   $\pm$   1.04    &	    22.12\\   
\object{HD\,113679}    	&	5768	$\pm$	    28	&	4.26	$\pm$	  0.02	&	1.08	$\pm$	  0.04	&	-0.61 $\pm$   0.02   &        252,33	&	0.83   $\pm$   0.03   &       0.83   $\pm$   0.03   &	    8.75    $\pm$   1.38    &	     9.45\\   
\object{HD\,119949}    	&	6359	$\pm$	    36	&	4.47	$\pm$	  0.04	&	1.65	$\pm$	  0.06	&	-0.41 $\pm$   0.02   &        226,35	&	1.07   $\pm$   0.04   &       1.03   $\pm$   0.03   &	    11.96   $\pm$   0.86    &	    17.86\\   
\object{HD\,121004}    	&	5687	$\pm$	    26	&	4.48	$\pm$	  0.03	&	0.76	$\pm$	  0.05	&	-0.71 $\pm$   0.02   &        248,32	&	0.79   $\pm$   0.03   &       0.79   $\pm$   0.03   &	    16.70   $\pm$   1.24    &	    17.54\\   
\object{HD\,123517}    	&	6082	$\pm$	    29	&	4.08	$\pm$	  0.05	&	1.53	$\pm$	  0.03	&	 0.09 $\pm$   0.02   &        252,36	&	       -	      &       1.21   $\pm$   0.08   &		    -		    &	     5.80\\   
\object{HD\,124785}    	&	5867	$\pm$	    21	&	4.20	$\pm$	  0.03	&	1.29	$\pm$	  0.03	&	-0.56 $\pm$   0.02   &        230,34	&	1.01   $\pm$   0.07   &       0.87   $\pm$   0.02   &	    8.09    $\pm$   1.00    &	    13.24\\   
\object{HD\,126681}    	&	5570	$\pm$	    34	&	4.70	$\pm$	  0.03	&	0.82	$\pm$	  0.10	&	-1.15 $\pm$   0.03   &        207,29	&	0.72   $\pm$   0.02   &       0.71   $\pm$   0.02   &	    21.04   $\pm$   1.12    &	    22.31\\   
\object{HD\,126793}    	&	5904	$\pm$	    33	&	4.43	$\pm$	  0.03	&	1.22	$\pm$	  0.06	&	-0.71 $\pm$   0.02   &        234,34	&	0.82   $\pm$   0.01   &       0.83   $\pm$   0.03   &	    18.53   $\pm$   0.97    &	    21.51\\   
\object{HD\,126803}    	&	5470	$\pm$	    18	&	4.48	$\pm$	  0.04	&	0.56	$\pm$	  0.05	&	-0.61 $\pm$   0.02   &        255,33	&	0.78   $\pm$   0.02   &       0.78   $\pm$   0.02   &	    19.18   $\pm$   1.21    &	    20.36\\   
\object{HD\,128340}    	&	6259	$\pm$	    40	&	4.64	$\pm$	  0.02	&	1.42	$\pm$	  0.08	&	-0.55 $\pm$   0.03   &        221,33	&	0.95   $\pm$   0.03   &       0.95   $\pm$   0.03   &	    11.62   $\pm$   0.95    &	    16.37\\   
\object{HD\,129229}    	&	5872	$\pm$	    21	&	3.89	$\pm$	  0.04	&	1.37	$\pm$	  0.03	&	-0.42 $\pm$   0.02   &        248,35	&	1.13   $\pm$   0.12   &       1.03   $\pm$   0.06   &	    7.10    $\pm$   1.35    &	     9.57\\   
\object{HD\,131653}    	&	5324	$\pm$	    26	&	4.54	$\pm$	  0.04	&	0.35	$\pm$	  0.09	&	-0.66 $\pm$   0.02   &        259,33	&	0.73   $\pm$   0.02   &       0.74   $\pm$   0.02   &	    20.17   $\pm$   1.16    &	    18.42\\   
\object{HD\,134088}    	&	5675	$\pm$	    22	&	4.46	$\pm$	  0.03	&	0.86	$\pm$	  0.04	&	-0.75 $\pm$   0.02   &        241,33	&	0.78   $\pm$   0.01   &       0.79   $\pm$   0.02   &	    26.62   $\pm$   0.86    &	    27.68\\   
\object{HD\,134113}    	&	5782	$\pm$	    22	&	4.25	$\pm$	  0.03	&	1.27	$\pm$	  0.04	&	-0.74 $\pm$   0.02   &        233,33	&	0.88   $\pm$   0.02   &       0.81   $\pm$   0.01   &	    13.91   $\pm$   1.21    &	    18.13\\   
\object{HD\,134440}    	&	4987	$\pm$	    48	&	4.80	$\pm$	  0.08	&	1.03	$\pm$	  0.20	&	-1.32 $\pm$   0.03   &        226,23	&	       -	      & 	     -  	    &	    35.14   $\pm$   1.48    &	    -	 \\   
\object{HD\,144589}    	&	6372	$\pm$	    37	&	4.28	$\pm$	  0.03	&	1.72	$\pm$	  0.05	&	-0.05 $\pm$   0.03   &        248,34	&	       -	      &       1.21   $\pm$   0.05   &		    -		    &	     5.80\\   
\object{HD\,145344}    	&	6143	$\pm$	    41	&	4.39	$\pm$	  0.04	&	1.48	$\pm$	  0.08	&	-0.68 $\pm$   0.03   &        219,34	&	0.91   $\pm$   0.02   &       0.88   $\pm$   0.03   &	    11.84   $\pm$   0.58    &	    16.91\\   
\object{HD\,145417}    	&	5006	$\pm$	    53	&	4.82	$\pm$	  0.12	&	0.65	$\pm$	  0.24	&	-1.23 $\pm$   0.04   &        147,15	&	       -	      & 	     -  	    &	    72.01   $\pm$   0.68    &	    -	 \\   
\object{HD\,147518}    	&	5626	$\pm$	    30	&	4.40	$\pm$	  0.03	&	0.67	$\pm$	  0.06	&	-0.63 $\pm$   0.02   &        257,34	&	0.80   $\pm$   0.02   &       0.80   $\pm$   0.03   &	    13.25   $\pm$   1.34    &	    14.25\\   
\object{HD\,148211}    	&	5948	$\pm$	    22	&	4.36	$\pm$	  0.02	&	1.40	$\pm$	  0.04	&	-0.62 $\pm$   0.02   &        233,33	&	0.87   $\pm$   0.01   &       0.85   $\pm$   0.02   &	    19.32   $\pm$   0.74    &	    24.34\\   
\object{HD\,148816}    	&	5908	$\pm$	    25	&	4.39	$\pm$	  0.02	&	1.36	$\pm$	  0.05	&	-0.71 $\pm$   0.02   &        230,34	&	0.86   $\pm$   0.00   &       0.83   $\pm$   0.02   &	    23.41   $\pm$   0.79    &	    31.54\\   
\object{HD\,149747}    	&	5823	$\pm$	    35	&	3.95	$\pm$	  0.04	&	1.28	$\pm$	  0.05	&	-0.34 $\pm$   0.03   &        259,34	&	       -	      &       0.99   $\pm$   0.06   &		    -		    &	     7.43\\   
\object{HD\,150177}    	&	6216	$\pm$	    28	&	4.18	$\pm$	  0.03	&	1.76	$\pm$	  0.06	&	-0.58 $\pm$   0.02   &        206,33	&	1.02   $\pm$   0.02   &       0.94   $\pm$   0.03   &	    24.96   $\pm$   0.63    &	    31.81\\   
\object{HD\,167300}    	&	5837	$\pm$	    20	&	4.30	$\pm$	  0.03	&	1.05	$\pm$	  0.03	&	-0.45 $\pm$   0.01   &        253,34	&	0.94   $\pm$   0.04   &       0.87   $\pm$   0.03   &	    8.97    $\pm$   1.24    &	    11.87\\   
\object{HD\,171028}    	&	5671	$\pm$	    16	&	3.84	$\pm$	  0.03	&	1.24	$\pm$	  0.02	&	-0.48 $\pm$   0.01   &        253,33	&	       -	      &       1.01   $\pm$   0.06   &		    -		    &	    10.38\\   
\object{HD\,171587}    	&	5412	$\pm$	    15	&	4.59	$\pm$	  0.02	&	0.76	$\pm$	  0.04	&	-0.64 $\pm$   0.01   &        256,33	&	0.77   $\pm$   0.02   &       0.76   $\pm$   0.02   &	    24.15   $\pm$   0.98    &	    29.14\\   
\object{HD\,175179}    	&	5764	$\pm$	    28	&	4.46	$\pm$	  0.03	&	0.88	$\pm$	  0.06	&	-0.66 $\pm$   0.02   &        241,33	&	0.81   $\pm$   0.03   &       0.81   $\pm$   0.03   &	    14.59   $\pm$   1.29    &	    16.14\\   
\object{HD\,175607}    	&	5392	$\pm$	    17	&	4.51	$\pm$	  0.03	&	0.60	$\pm$	  0.04	&	-0.62 $\pm$   0.01   &        258,33	&	0.77   $\pm$   0.02   &       0.77   $\pm$   0.02   &	    22.09   $\pm$   1.01    &	    25.67\\   
\object{HD\,181720}    	&	5792	$\pm$	    17	&	4.25	$\pm$	  0.02	&	1.16	$\pm$	  0.02	&	-0.53 $\pm$   0.01   &        253,34	&	0.92   $\pm$   0.01   &       0.86   $\pm$   0.02   &	    17.22   $\pm$   1.16    &	    21.23\\   
\object{HD\,190984}     &	6007	$\pm$	    25  &	4.02	$\pm$	  0.03  &	1.58	$\pm$	  0.03  &	-0.49 $\pm$   0.02   &        233,35	&	1.16   $\pm$   0.12   &       0.96   $\pm$   0.04   &	    5.46    $\pm$   1.11    &	     9.25\\   
\object{HD\,193901}     &	5611	$\pm$	    34  &	4.41	$\pm$	  0.05  &	0.54	$\pm$	  0.11  &	-1.07 $\pm$   0.03   &        222,31	&	0.74   $\pm$   0.02   &       0.74   $\pm$   0.02   &	    22.78   $\pm$   1.00    &	    20.59\\   
\object{HD\,195633}     &	6154	$\pm$	    37  &	4.25	$\pm$	  0.05  &	1.47	$\pm$	  0.06  &	-0.51 $\pm$   0.03   &        230,34	&	0.98   $\pm$   0.03   &       0.93   $\pm$   0.03   &	    10.07   $\pm$   0.84    &	    12.84\\   
\object{HD\,196892}     &	6072	$\pm$	    56  &	4.50	$\pm$	  0.03  &	1.21	$\pm$	  0.12  &	-0.89 $\pm$   0.03   &        204,32	&	0.83   $\pm$   0.01   &       0.83   $\pm$   0.03   &	    16.15   $\pm$   0.93    &	    21.48\\   
\object{HD\,197083}     &	5735	$\pm$	    16  &	4.50	$\pm$	  0.02  &	0.90	$\pm$	  0.02  &	-0.45 $\pm$   0.01   &        255,33	&	0.85   $\pm$   0.03   &       0.85   $\pm$   0.03   &	    13.75   $\pm$   1.14    &	    15.77\\   
\object{HD\,197197}     &	5812	$\pm$	    16  &	4.20	$\pm$	  0.02  &	1.25	$\pm$	  0.02  &	-0.46 $\pm$   0.01   &        254,33	&	0.93   $\pm$   0.01   &       0.89   $\pm$   0.02   &	    14.50   $\pm$   0.92    &	    17.50\\   
\object{HD\,197536}     &	6105	$\pm$	    24  &	4.39	$\pm$	  0.03  &	1.34	$\pm$	  0.04  &	-0.41 $\pm$   0.02   &        243,34	&	0.96   $\pm$   0.03   &       0.94   $\pm$   0.04   &	    14.15   $\pm$   0.93    &	    17.79\\   
\object{HD\,199288}     &	5746	$\pm$	    22  &	4.46	$\pm$	  0.03  &	0.93	$\pm$	  0.04  &	-0.63 $\pm$   0.02   &        251,33	&	0.82   $\pm$   0.00   &       0.82   $\pm$   0.03   &	    45.17   $\pm$   0.46    &	    51.99\\   
\object{HD\,199289}     &	5928	$\pm$	    37  &	4.64	$\pm$	  0.03  &	1.30	$\pm$	  0.10  &	-0.98 $\pm$   0.03   &        207,31	&	0.79   $\pm$   0.01   &       0.79   $\pm$   0.03   &	    18.95   $\pm$   0.76    &	    26.79\\   
\object{HD\,199604}     &	5817	$\pm$	    22  &	4.34	$\pm$	  0.03  &	1.04	$\pm$	  0.04  &	-0.62 $\pm$   0.02   &        243,33	&	0.83   $\pm$   0.01   &       0.83   $\pm$   0.02   &	    14.82   $\pm$   0.96    &	    17.02\\   
\object{HD\,199847}     &	5763	$\pm$	    20  &	4.22	$\pm$	  0.02  &	1.04	$\pm$	  0.03  &	-0.54 $\pm$   0.02   &        254,33	&	0.85   $\pm$   0.03   &       0.85   $\pm$   0.02   &	    12.73   $\pm$   1.16    &	    13.42\\   
\object{HD\,206998}     &	5822	$\pm$	    26  &	4.24	$\pm$	  0.03  &	1.13	$\pm$	  0.04  &	-0.69 $\pm$   0.02   &        240,34	&	0.89   $\pm$   0.02   &       0.82   $\pm$   0.02   &	    11.32   $\pm$   1.08    &	    14.36\\   
\object{HD\,207190}     &	6178	$\pm$	    26  &	4.33	$\pm$	  0.03  &	1.51	$\pm$	  0.04  &	-0.42 $\pm$   0.02   &        232,35	&	0.99   $\pm$   0.03   &       0.96   $\pm$   0.03   &	    16.79   $\pm$   0.75    &	    20.42\\   
\object{HD\,207869}     &	5527	$\pm$	    21  &	4.50	$\pm$	  0.05  &	0.73	$\pm$	  0.05  &	-0.45 $\pm$   0.02   &        259,33	&	0.81   $\pm$   0.03   &       0.82   $\pm$   0.03   &	    21.55   $\pm$   1.10    &	    19.58\\   
\object{HD\,210752}     &	5951	$\pm$	    21  &	4.53	$\pm$	  0.03  &	1.20	$\pm$	  0.04  &	-0.58 $\pm$   0.02   &        235,34	&	0.86   $\pm$   0.02   &       0.87   $\pm$   0.04   &	    27.64   $\pm$   0.68    &	    32.91\\   
\object{HD\,215257}     &	6052	$\pm$	    26  &	4.46	$\pm$	  0.02  &	1.40	$\pm$	  0.05  &	-0.63 $\pm$   0.02   &        226,34	&	0.87   $\pm$   0.01   &       0.87   $\pm$   0.04   &	    23.27   $\pm$   0.67    &	    29.47\\   
\object{HD\,218504}    	&	5962	$\pm$	    29	&	4.34	$\pm$	  0.03	&	1.21	$\pm$	  0.05	&	-0.55 $\pm$   0.02   &        240,33	&	0.90   $\pm$   0.01   &       0.87   $\pm$   0.02   &	    15.27   $\pm$   0.78    &	    19.26\\  
\object{HD\,221580}    	&	5322	$\pm$	    24	&	2.68	$\pm$	  0.04	&	1.97	$\pm$	  0.04	&	-1.13 $\pm$   0.02   &        216,34	&	0.98   $\pm$   0.27   &       1.10   $\pm$   0.25   &	    0.76    $\pm$   1.28    &	     2.08\\   
\object{HD\,223854}    	&	6080	$\pm$	    30	&	4.08	$\pm$	  0.03	&	1.60	$\pm$	  0.05	&	-0.54 $\pm$   0.02   &        228,34	&	0.99   $\pm$   0.03   &       0.94   $\pm$   0.03   &	    11.98   $\pm$   0.81    &	    13.62\\   
\object{HD\,224347}    	&	6092	$\pm$	    24	&	4.27	$\pm$	  0.03	&	1.31	$\pm$	  0.04	&	-0.42 $\pm$   0.02   &        237,35	&	0.95   $\pm$   0.03   &       0.94   $\pm$   0.03   &	    12.39   $\pm$   0.89    &	    13.63\\   
\object{HD\,224817}    	&	5894	$\pm$	    22	&	4.36	$\pm$	  0.02	&	1.13	$\pm$	  0.04	&	-0.53 $\pm$   0.02   &        247,33	&	0.91   $\pm$   0.02   &       0.86   $\pm$   0.03   &	    13.68   $\pm$   1.22    &	    17.82\\   
\object{BD+062932} 	&	5272	$\pm$	    37	&	4.43	$\pm$	  0.05	&	0.06	$\pm$	  0.62	&	-0.84 $\pm$   0.04   &        255,33	&	       -	      & 	     -  	    &	    12.86   $\pm$   2.76    &	    -	 \\   
\object{BD+063077}	&	6136	$\pm$	   171	&	4.95	$\pm$	  0.15	&	1.07	$\pm$	  0.38	&	-0.36 $\pm$   0.12   &        247,35	&	       -	      & 	     -  	    &	    7.03    $\pm$   2.27    &	    -	 \\   
\object{BD+083095}	&	5728	$\pm$	    41	&	4.12	$\pm$	  0.04	&	0.85	$\pm$	  0.08	&	-0.77 $\pm$   0.03   &        239,34	&	       -	      & 	     -  	    &	    4.82    $\pm$   2.19    &	    -	 \\   
\object{BD-084501} 	&	6216	$\pm$	    84	&	4.81	$\pm$	  0.07	&	2.36	$\pm$	  0.73	&	-1.39 $\pm$   0.06   &         74,13	&	0.80   $\pm$   0.03   &       0.80   $\pm$   0.03   &	    9.59    $\pm$   2.21    &	    10.01\\   
\object{CD-2310879}	&	6788	$\pm$	    43	&	4.67	$\pm$	  0.04	&	1.82	$\pm$	  0.09	&	-0.24 $\pm$   0.02   &        211,33	&	1.20   $\pm$   0.03   &       1.20   $\pm$   0.03   &	    7.41    $\pm$   1.69    &	     7.39\\   
\object{CD-436810} 	&	6011	$\pm$	    28	&	4.41	$\pm$	  0.02	&	1.09	$\pm$	  0.04	&	-0.44 $\pm$   0.02   &        252,35	&	0.91   $\pm$   0.03   &       0.91   $\pm$   0.04   &	    9.27    $\pm$   1.29    &	     9.14\\   
\object{CD-4512460} 	&	5960	$\pm$	    69	&	4.42	$\pm$	  0.05	&	0.75	$\pm$	  0.17	&	-0.86 $\pm$   0.05   &        216,33	&		-	      &       0.81   $\pm$   0.03   &		    -		    &	     5.90\\   
\object{CD-452997} 	&	5312	$\pm$	    34	&	4.39	$\pm$	  0.06	&	0.24	$\pm$	  0.19	&	-0.84 $\pm$   0.03   &        247,34	&		-	      &       0.72   $\pm$   0.02   &		    -		    &	     9.03\\   
\object{CD-571633} 	&	5975	$\pm$	    41	&	4.46	$\pm$	  0.03	&	1.14	$\pm$	  0.08	&	-0.85 $\pm$   0.03   &        216,33	&		-	      &       0.82   $\pm$   0.03   &	    9.91    $\pm$   0.88    &	    11.78\\   

\end{longtable}
\tablefoot{$\log{g}_{spec}$ the spectroscopic surface gravity; $\xi_{\mathrm{t}}$ is the microturbulance speed; N(\ion{Fe}{i},\ion{Fe}{ii}) is the number of lines used in the spectroscopic 
analysis; $Mass_{hip}$ is the mass determined directly from the Padova webinterface using the Hipparcus parallax; $Mass_{est}$ is the mass determined using the iterative procedure to compute new paralaxe 
values (See text for more details); $\pi_{phip}$ the Hipparcus parallax; and $\pi_{pest}$ is the estimated parallax from the iterative procedure.}
\end{scriptsize}
}

\longtab{3}{
\begin{scriptsize}
\begin{longtable}{cccccccccccc}
\caption{\label{tabphoto} List of the photometry used for the use of the IFRM calibration }\\
\hline\hline
 & & & & & & & & & & & \\
star & Umag$^{(1)}$ & Bmag$^{(2)}$ & Vmag$^{(2)}$ & Imag$^{(2)}$ & V-I$^{(3)}$ & Jmag$^{(4)}$ & Hmag$^{(4)}$ & Kmag$^{(4)}$ & eJmag$^{(4)}$ & eHmag$^{(4)}$ & eKmag$^{(4)}$ \\
\hline\hline
\endfirsthead
\caption{continued.}\\
\hline\hline
star & Umag & Bmag & Vmag & Imag & V-I & Jmag & Hmag & Kmag & eJmag & eHmag & eKmag \\
\hline
\endhead
\hline
\endfoot

\object{HD\,967}	&	9.03	&	8.959	&	8.389	&	8.000	&	0.710	&	7.136	&	6.816	&	6.722	&	0.021	&	0.055	&	0.024	\\
\object{HD\,11397}	&	9.75	&	9.574	&	8.944	&	8.520	&	0.750	&	7.686	&	7.318	&	7.270	&	0.023	&	0.033	&	0.023	\\
\object{HD\,16784}	&	8.54	&	8.548	&	8.036	&	7.700	&	0.640	&	6.861	&	6.530	&	6.477	&	0.020	&	0.033	&	0.017	\\
\object{HD\,17548}	&	-	&	8.631	&	8.185	&	7.900	&	0.600	&	7.097	&	6.802	&	6.765	&	0.021	&	0.051	&	0.024	\\
\object{HD\,17865}	&	-	&	8.696	&	8.171	&	7.810	&	0.660	&	7.086	&	6.786	&	6.695	&	0.032	&	0.040	&	0.021	\\
\object{HD\,22879}	&	-	&	-	&	6.68*	&	-	&	0.660	&	5.588	&	5.301	&	5.179	&	0.019	&	0.029	&	0.021	\\
\object{HD\,25704}	&	-	&	7.188	&	6.691	&	6.350	&	0.640	&	6.977	&	6.650	&	6.556	&	0.026	&	0.020	&	0.029	\\
\object{HD\,31128}	&	9.43	&	9.532	&	9.157	&	8.910	&	0.570	&	8.032	&	7.800	&	7.738	&	0.023	&	0.027	&	0.018	\\
\object{HD\,38510}	&	8.66	&	8.703	&	8.226	&	7.910	&	0.610	&	7.114	&	6.778	&	6.742	&	0.023	&	0.046	&	0.023	\\
\object{HD\,40865}	&	9.27	&	9.193	&	8.595	&	8.200	&	0.700	&	7.402	&	7.130	&	7.049	&	0.019	&	0.031	&	0.021	\\
\object{HD\,51754}	&	9.58	&	9.560	&	9.010	&	8.640	&	0.650	&	7.867	&	7.602	&	7.530	&	0.019	&	0.031	&	0.024	\\
\object{HD\,56274}	&	8.34	&	8.313	&	7.758	&	7.390	&	0.680	&	6.593	&	6.271	&	6.203	&	0.020	&	0.044	&	0.026	\\
\object{HD\,59984}	&	-	&	6.402	&	5.925	&	5.610	&	0.690	&	5.092	&	4.580	&	4.480	&	0.266	&	0.076	&	0.016	\\
\object{HD\,61902}	&	8.61	&	8.685	&	8.242	&	7.940	&	0.560	&	7.222	&	6.960	&	6.888	&	0.021	&	0.046	&	0.021	\\
\object{HD\,62849}	&	-	&	10.760	&	9.781	&	9.170	&	-	&	8.004	&	7.544	&	7.435	&	0.020	&	0.036	&	0.024	\\
\object{HD\,68089}	&	-	&	10.157	&	9.575	&	9.190	&	0.670	&	8.344	&	8.020	&	7.977	&	0.024	&	0.059	&	0.029	\\
\object{HD\,68284}	&	-	&	8.249	&	7.757	&	7.430	&	0.660	&	6.649	&	6.338	&	6.269	&	0.018	&	0.027	&	0.027	\\
\object{HD\,69611}	&	8.25	&	8.279	&	7.739	&	7.370	&	0.660	&	6.594	&	6.275	&	6.213	&	0.019	&	0.046	&	0.021	\\
\object{HD\,75745}	&	-	&	9.986	&	9.445	&	9.080	&	-	&	8.329	&	8.029	&	7.935	&	0.029	&	0.044	&	0.024	\\
\object{HD\,77110}	&	-	&	9.421	&	8.862	&	8.480	&	0.690	&	7.667	&	7.358	&	7.268	&	0.024	&	0.040	&	0.023	\\
\object{HD\,78747}	&	8.24	&	8.248	&	7.717	&	7.360	&	0.650	&	6.542	&	6.214	&	6.179	&	0.024	&	0.036	&	0.024	\\
\object{HD\,79601}	&	-	&	8.559	&	8.010	&	7.640	&	0.670	&	6.872	&	6.563	&	6.494	&	0.021	&	0.026	&	0.021	\\
\object{HD\,88474}	&	-	&	8.983	&	8.476	&	8.140	&	0.680	&	7.311	&	7.035	&	6.941	&	0.020	&	0.031	&	0.021	\\
\object{HD\,88725}	&	8.35	&	8.309	&	7.748	&	7.370	&	0.680	&	6.543	&	6.242	&	6.153	&	0.020	&	0.034	&	0.024	\\
\object{HD\,90422}	&	-	&	8.730	&	8.253	&	7.930	&	0.590	&	7.176	&	6.904	&	6.853	&	0.027	&	0.026	&	0.016	\\
\object{HD\,91345}	&	9.5	&	9.580	&	9.066	&	8.730	&	0.630	&	7.877	&	7.571	&	7.518	&	0.021	&	0.029	&	0.036	\\
\object{HD\,94444}	&	8.52	&	8.550	&	8.095	&	7.800	&	0.620	&	7.024	&	6.778	&	6.665	&	0.027	&	0.042	&	0.021	\\
\object{HD\,95860}	&	-	&	10.272	&	9.745	&	9.400	&	-	&	8.655	&	8.387	&	8.312	&	0.018	&	0.036	&	0.031	\\
\object{HD\,97320}	&	8.48	&	8.593	&	8.188	&	7.920	&	0.560	&	7.137	&	6.868	&	6.790	&	0.023	&	0.031	&	0.017	\\
\object{HD\,97783}	&	9.65	&	9.617	&	9.048	&	8.660	&	0.660	&	7.888	&	7.615	&	7.506	&	0.024	&	0.053	&	0.033	\\
\object{HD\,102200}	&	9	&	9.138	&	8.746	&	8.490	&	0.500	&	7.688	&	7.449	&	7.383	&	0.020	&	0.042	&	0.024	\\
\object{HD\,104800}	&	9.76	&	9.773	&	9.224	&	8.850	&	0.660	&	8.084	&	7.765	&	7.669	&	0.024	&	0.042	&	0.024	\\
\object{HD\,105004}	&	-	&	10.935	&	10.403	&	10.030	&	0.630	&	9.229	&	8.949	&	8.865	&	0.029	&	0.051	&	0.023	\\
\object{HD\,107094}	&	-	&	9.757	&	9.139	&	8.720	&	0.720	&	7.829	&	7.531	&	7.407	&	0.021	&	0.049	&	0.024	\\
\object{HD\,108564}	&	-	&	10.380	&	9.460	&	8.880	&	1.010	&	-	&	-	&	-	&	-	&	-	&	-	\\
\object{HD\,109310}	&	-	&	8.868	&	8.362	&	8.020	&	0.640	&	7.280	&	7.031	&	6.913	&	0.023	&	0.038	&	0.023	\\
\object{HD\,109684}	&	-	&	9.246	&	8.735	&	8.400	&	0.660	&	7.649	&	7.379	&	7.319	&	0.027	&	0.040	&	0.027	\\
\object{HD\,111515}	&	8.97	&	8.777	&	8.124	&	7.680	&	0.740	&	6.812	&	6.493	&	6.358	&	0.023	&	0.053	&	0.020	\\
\object{HD\,111777}	&	9.07	&	9.060	&	8.501	&	8.120	&	0.680	&	7.274	&	6.956	&	6.896	&	0.023	&	0.049	&	0.021	\\
\object{HD\,113679}	&	10.3	&	10.265	&	9.754	&	9.400	&	0.680	&	8.466	&	8.228	&	8.108	&	0.018	&	0.046	&	0.029	\\
\object{HD\,119949}	&	-	&	8.579	&	8.166	&	7.900	&	0.600	&	7.093	&	6.810	&	6.792	&	0.027	&	0.036	&	0.023	\\
\object{HD\,121004}	&	9.59	&	9.578	&	9.034	&	8.660	&	0.670	&	7.817	&	7.533	&	7.426	&	0.020	&	0.061	&	0.026	\\
\object{HD\,123517}	&	-	&	10.210	&	9.563	&	9.120	&	-	&	8.232	&	7.934	&	7.874	&	0.032	&	0.036	&	0.027	\\
\object{HD\,124785}	&	-	&	9.186	&	8.680	&	8.340	&	0.650	&	7.496	&	7.202	&	7.147	&	0.021	&	0.023	&	0.026	\\
\object{HD\,126681}	&	9.8	&	9.847	&	9.298	&	8.930	&	0.680	&	8.044	&	7.709	&	7.631	&	0.023	&	0.040	&	0.024	\\
\object{HD\,126793}	&	-	&	8.731	&	8.246	&	7.930	&	0.600	&	7.082	&	6.818	&	6.722	&	0.020	&	0.042	&	0.020	\\
\object{HD\,126803}	&	9.71	&	9.583	&	8.904	&	8.450	&	0.750	&	7.689	&	7.306	&	7.240	&	0.030	&	0.038	&	0.021	\\
\object{HD\,128340}	&	-	&	9.323	&	8.892	&	8.600	&	0.610	&	7.840	&	7.596	&	7.531	&	0.021	&	0.024	&	0.020	\\
\object{HD\,128575}	&	-	&	9.036	&	8.440	&	8.030	&	0.700	&	7.268	&	7.003	&	6.852	&	0.024	&	0.042	&	0.018	\\
\object{HD\,129229}	&	-	&	8.955	&	8.418	&	8.050	&	0.700	&	7.203	&	6.896	&	6.822	&	0.020	&	0.040	&	0.020	\\
\object{HD\,131653}	&	10.4	&	10.186	&	9.567	&	9.150	&	0.760	&	8.154	&	7.802	&	7.680	&	0.035	&	0.051	&	0.033	\\
\object{HD\,134088}	&	8.53	&	8.554	&	8.021	&	7.660	&	0.670	&	6.807	&	6.490	&	6.434	&	0.029	&	0.047	&	0.033	\\
\object{HD\,134113}	&	8.79	&	8.793	&	8.285	&	7.950	&	0.650	&	7.093	&	6.761	&	6.695	&	0.030	&	0.031	&	0.017	\\
\object{HD\,134440}	&	10.63	&	10.316	&	9.495	&	8.970	&	0.860	&	-	&	-	&	-	&	-	&	-	&	-	\\
\object{HD\,136269}	&	-	&	10.401	&	9.716	&	9.270	&	-	&	8.233	&	7.810	&	7.688	&	0.030	&	0.057	&	0.040	\\
\object{HD\,144589}	&	-	&	10.315	&	9.812	&	9.470	&	-	&	8.705	&	8.491	&	8.339	&	0.027	&	0.059	&	0.018	\\
\object{HD\,145344}	&	-	&	8.842	&	8.377	&	8.070	&	0.630	&	7.340	&	7.101	&	7.048	&	0.027	&	0.049	&	0.023	\\
\object{HD\,145417}	&	8.64	&	8.338	&	7.543	&	-	&	0.940	&	-	&	-	&	-	&	-	&	-	&	-	\\
\object{HD\,147518}	&	-	&	9.972	&	9.386	&	9.000	&	0.680	&	8.106	&	7.777	&	7.764	&	0.021	&	0.038	&	0.026	\\
\object{HD\,148211}	&	8.18	&	8.197	&	7.708	&	7.380	&	0.620	&	6.569	&	6.284	&	6.202	&	0.024	&	0.040	&	0.021	\\
\object{HD\,148816}	&	7.75	&	7.798	&	7.287	&	6.950	&	0.650	&	6.159	&	5.862	&	5.809	&	0.027	&	0.026	&	0.020	\\
\object{HD\,149747}	&	-	&	9.772	&	9.174	&	8.860	&	-	&	7.947	&	7.643	&	7.544	&	0.034	&	0.021	&	0.036	\\
\object{HD\,150177}	&	-	&	6.762	&	6.341	&	6.060	&	0.560	&	5.353	&	5.064	&	4.977	&	0.037	&	0.040	&	0.018	\\
\object{HD\,161265}	&	-	&	10.255	&	9.739	&	9.390	&	-	&	8.579	&	8.404	&	8.302	&	0.030	&	0.023	&	0.025	\\
\object{HD\,164500}	&	-	&	10.327	&	9.643	&	9.180	&	0.830	&	8.192	&	7.782	&	7.714	&	0.024	&	0.044	&	0.017	\\
\object{HD\,167300}	&	9.78	&	9.746	&	9.258	&	8.930	&	0.660	&	8.013	&	7.757	&	7.628	&	0.021	&	0.059	&	0.018	\\
\object{HD\,171028}	&	-	&	8.905	&	8.301	&	7.890	&	-	&	6.990	&	6.663	&	6.604	&	0.027	&	0.036	&	0.027	\\
\object{HD\,171587}	&	9.25	&	9.211	&	8.544	&	8.090	&	0.750	&	7.197	&	6.839	&	6.754	&	0.026	&	0.029	&	0.027	\\
\object{HD\,175179}	&	9.6	&	9.609	&	9.092	&	8.740	&	0.660	&	7.929	&	7.630	&	7.543	&	0.023	&	0.046	&	0.027	\\
\object{HD\,175607}	&	-	&	9.240	&	8.606	&	8.190	&	0.760	&	7.280	&	6.928	&	6.881	&	0.020	&	0.026	&	0.057	\\
\object{HD\,181720}	&	8.41	&	8.404	&	7.858	&	7.490	&	0.670	&	6.652	&	6.346	&	6.294	&	0.019	&	0.029	&	0.034	\\
\object{HD\,187151}	&	-	&	9.243	&	8.624	&	8.200	&	-	&	7.359	&	7.114	&	6.982	&	0.021	&	0.049	&	0.017	\\
\object{HD\,190984}	&	9.32	&	9.285	&	8.770	&	8.420	&	0.650	&	7.671	&	7.389	&	7.319	&	0.023	&	0.024	&	0.016	\\
\object{HD\,193901}	&	9.07	&	9.137	&	8.658	&	8.340	&	0.630	&	-	&	-	&	-	&	-	&	-	&	-	\\
\object{HD\,195633}	&	9.01	&	9.007	&	8.506	&	8.170	&	0.620	&	7.442	&	7.169	&	7.104	&	0.024	&	0.024	&	0.021	\\
\object{HD\,196892}	&	8.59	&	8.700	&	8.261	&	7.970	&	0.560	&	7.182	&	6.907	&	6.824	&	0.026	&	0.047	&	0.017	\\
\object{HD\,197083}	&	-	&	9.779	&	9.198	&	8.810	&	0.710	&	8.035	&	7.755	&	7.636	&	0.027	&	0.044	&	0.038	\\
\object{HD\,197197}	&	-	&	8.657	&	8.075	&	7.690	&	0.670	&	6.908	&	6.624	&	6.519	&	0.027	&	0.046	&	0.021	\\
\object{HD\,197536}	&	-	&	8.696	&	8.208	&	7.880	&	0.620	&	7.127	&	6.853	&	6.808	&	0.020	&	0.024	&	0.023	\\
\object{HD\,197890}	&	-	&	10.377	&	9.454	&	8.870	&	0.930	&	7.513	&	6.930	&	6.794	&	0.021	&	0.029	&	0.026	\\
\object{HD\,199288}	&	-	&	7.055	&	6.518	&	6.150	&	0.680	&	5.446	&	5.153	&	5.018	&	0.017	&	0.055	&	0.017	\\
\object{HD\,199289}	&	8.68	&	8.764	&	8.284	&	7.960	&	0.610	&	7.178	&	6.920	&	6.841	&	0.020	&	0.046	&	0.024	\\
\object{HD\,199604}	&	-	&	9.114	&	8.617	&	8.290	&	0.640	&	7.413	&	7.161	&	7.099	&	0.023	&	0.040	&	0.026	\\
\object{HD\,199847}	&	-	&	9.375	&	8.823	&	8.440	&	0.680	&	7.634	&	7.349	&	7.257	&	0.029	&	0.061	&	0.023	\\
\object{HD\,206998}	&	-	&	9.204	&	8.680	&	8.330	&	0.630	&	7.556	&	7.212	&	7.198	&	0.029	&	0.036	&	0.023	\\
\object{HD\,207190}	&	-	&	8.128	&	7.672	&	7.360	&	0.600	&	6.691	&	6.424	&	6.352	&	0.027	&	0.069	&	0.027	\\
\object{HD\,207869}	&	-	&	9.584	&	8.974	&	8.560	&	0.730	&	7.658	&	7.328	&	7.228	&	0.023	&	0.027	&	0.021	\\
\object{HD\,210752}	&	7.84	&	7.940	&	7.445	&	7.120	&	0.610	&	6.365	&	6.070	&	6.052	&	0.024	&	0.038	&	0.024	\\
\object{HD\,215257}	&	7.82	&	7.891	&	7.431	&	7.120	&	0.590	&	6.309	&	6.021	&	5.951	&	0.019	&	0.040	&	0.020	\\
\object{HD\,218504}	&	-	&	8.620	&	8.121	&	7.780	&	0.650	&	7.014	&	6.730	&	6.666	&	0.018	&	0.034	&	0.021	\\
\object{HD\,221580}	&	-	&	9.828	&	9.215	&	8.810	&	0.750	&	7.736	&	7.370	&	7.247	&	0.024	&	0.051	&	0.017	\\
\object{HD\,223854}	&	8.48	&	8.509	&	8.053	&	7.740	&	0.590	&	6.956	&	6.719	&	6.667	&	0.021	&	0.031	&	0.021	\\
\object{HD\,224347}	&	-	&	8.969	&	8.489	&	8.170	&	0.620	&	7.457	&	7.170	&	7.118	&	0.023	&	0.042	&	0.023	\\
\object{HD\,224817}	&	8.87	&	8.929	&	8.414	&	8.060	&	0.640	&	7.270	&	7.017	&	6.904	&	0.021	&	0.059	&	0.020	\\
\object{BD+062932}	&	-	&	11.306	&	10.515	&	-	&	0.760	&	9.030	&	8.577	&	8.634	&	0.023	&	0.029	&	0.027	\\
\object{BD+063077}	&	-	&	10.955	&	10.288	&	10.390	&	0.610	&	9.329	&	9.035	&	8.973	&	0.023	&	0.023	&	0.021	\\
\object{BD+083095}	&	10.55	&	10.544	&	10.098	&	9.810	&	0.660	&	8.750	&	8.410	&	8.375	&	0.021	&	0.046	&	0.019	\\
\object{BD-004234}	&	11.31	&	10.637	&	9.753	&	9.190	&	1.000	&	7.770	&	7.240	&	7.082	&	0.018	&	0.038	&	0.029	\\
\object{BD-032525}	&	-	&	10.076	&	9.677	&	9.680	&	0.550	&	8.561	&	8.276	&	8.205	&	0.026	&	0.044	&	0.026	\\
\object{BD-084501}	&	-	&	11.018	&	10.667	&	10.440	&	0.670	&	9.172	&	8.863	&	8.770	&	0.026	&	0.042	&	0.021	\\
\object{CD-436810}	&	-	&	10.299	&	9.828	&	9.510	&	0.640	&	8.740	&	8.495	&	8.395	&	0.024	&	0.038	&	0.020	\\
\object{CD-452997}	&	-	&	11.311	&	10.751	&	10.370	&	-	&	9.292	&	8.857	&	8.800	&	0.024	&	0.026	&	0.025	\\
\object{CD-571633}	&	-	&	9.992	&	9.478	&	9.440	&	0.560	&	8.499	&	8.219	&	8.089	&	0.021	&	0.049	&	0.036	\\

\end{longtable}
\tablefoot{(1) from the SKY2000 Catalog, Version 4 \citep[][]{Myers-2001}. (2) were obtained from the NOMAD 
catalog \citep[][]{Zacharias-2004}. (3) from the index V-I obtained in the Hipparcus catalog \citep[][]{Hipparcus-2007}. (4) were 
obtained from the 2MASS catalog \citep[][]{2MASS-2003}. *The Vmag from the star HD22879 was also taken from the Hipparcus catalog.}
\end{scriptsize}
}

\end{document}